\documentclass[journal=ancac3, manuscript=article]{achemso}

\usepackage{chemformula} 
\usepackage[T1]{fontenc} 

\usepackage{amssymb}
\usepackage{subfigure,amsfonts, latexsym}

\usepackage{epsfig,graphicx,subfigure,dcolumn,bm}

\author{Wei-Chi Chiu}
\email{chiu.w@northeastern.edu}
\affiliation{Department of Physics, Northeastern University, Boston, MA 02115, USA}
\alsoaffiliation{Quantum Materials and Sensing Institute, Northeastern University, Burlington, MA 01803, USA}

\author{Sougata Mardanya}
\affiliation{Department of Physics, National Cheng Kung University, Tainan 70101, Taiwan}
\altaffiliation{Current address: Department of Physics and Astronomy, Howard University, Washington, DC 20059, USA}

\author{Robert Markiewicz}
\affiliation{Department of Physics, Northeastern University, Boston, MA 02115, USA}
\alsoaffiliation{Quantum Materials and Sensing Institute, Northeastern University, Burlington, MA 01803, USA}

\author{Jouko Nieminen}
\affiliation{Computational Physics Laboratory, Tampere University, Tampere 33014, Finland}
\alsoaffiliation{Department of Physics, Northeastern University, Boston, MA 02115, USA}
\alsoaffiliation{Quantum Materials and Sensing Institute, Northeastern University, Burlington, MA 01803, USA}

\author{Bahadur Singh}
\affiliation{Department of Condensed Matter Physics and Materials Science, Tata Institute of Fundamental Research, Colaba, Mumbai 400005, India}

\author{Tugrul Hakioglu}
\affiliation{Energy Institute and Department of Physics, Istanbul Technical University, Maslak 34469, Istanbul, Turkey}
\alsoaffiliation{Department of Physics, University of Michigan, Ann Arbor, MI 48109}

\author{Amit Agarwal}
\affiliation{Department of Physics, Indian Institute of Technology Kanpur, Kanpur 208016, India}

\author{Tay-Rong Chang}
\affiliation{Department of Physics, National Cheng Kung University, Tainan 70101, Taiwan}
\alsoaffiliation{ Center for Quantum Frontiers of Research and Technology (QFort), Tainan 701, Taiwan}
\alsoaffiliation{Physics Division, National Center for Theoretical Sciences, Taipei 10617, Taiwan}

\author{Hsin Lin}
\affiliation{Institute of Physics, Academia Sinica, Taipei 115201, Taiwan}

\author{Arun Bansil}
\email{ar.bansil@northeastern.edu}
\affiliation{Department of Physics, Northeastern University, Boston, MA 02115, USA}
\alsoaffiliation{Quantum Materials and Sensing Institute, Northeastern University, Burlington, MA 01803, USA}

\title[An \textsf{achemso} demo]
  {Strain-Induced Charge Density Waves with Emergent Topological States in Monolayer NbSe$_{2}$}


\begin{document}

\clearpage
\begin{abstract}
Emergence of topological states in strongly correlated systems, particularly two-dimensional (2D) transition-metal dichalcogenides, offers a platform for manipulating electronic properties in quantum materials. However, a comprehensive understanding of the intricate interplay between correlations and topology remains elusive. Here we employ first-principles modeling to reveal two distinct 2$\times$2 charge density wave (CDW) phases in monolayer 1H-NbSe$_2$, which become energetically favorable over the conventional 3$\times$3 CDWs under modest biaxial tensile strain of about 1\%. These strain-induced CDW phases coexist with numerous topological states characterized by $\mathbb{Z}_2$ topology, high mirror Chern numbers, topological nodal lines, and higher-order topological states, which we have verified rigorously by computing the topological indices and the presence of robust edge states and localized corner states. Remarkably, these topological properties emerge because of the CDW rather than a pre-existing topology in the pristine phase. These results elucidate the interplay between correlations, topology, and geometry in 2D materials and indicate that strain‑induced correlation effects can be used to engineer topological states in materials with initially trivial topology. Our findings may be applied in electronics, spintronics, and other advanced quantum devices that require robust and tunable topological states.

\end{abstract}

\newpage
The field of two-dimensional (2D) materials~\cite{Butler:2013aa} continues to expand unabated, where the transition-metal dichalcogenides (TMDs) are at the forefront due to their exceptional electronic properties and potential for novel applications~\cite{Manzeli:2017aa, CHOI2017116}. Given their propensity to host a broad range of quantum states, including topological states~\cite{Qian1344, PhysRevMaterials.7.064001}, charge density waves~\cite{Xu:2021aa, Flicker:2015uy}, and superconductivity~\cite{Ugeda:2016aa, PhysRevB.93.180501}‚ the 2D TMDs provide a versatile platform for developing advanced devices in diverse fields such as optoelectronics~\cite{Wang:2012aa}, transistors~\cite{Radisavljevic:2011aa}, spintronics~\cite{Sierra:2021aa}, valleytronics~\cite{Schaibley:2016aa}, and Josephson junctions~\cite{Dvir:2018aa, PhysRevB.107.174524}. Moreover, the advent of TMD monolayers, known for their exceptional elasticity and resilience to large in-plane strains ($ \gtrsim10\%$)~\cite{Rold_n_2015} offer a fertile ground for engineering topological phases~\cite{Tang:2017aa, Fei:2017wy, Shi:2019aa} and manipulating many-body ground states~\cite{Zhou:2012aa, C4NR01486C, PhysRevB.97.081101}. 

Despite recent advances, a full understanding of the interplay between correlation effects and topological states in the TMDs remains a challenge and presents a new frontier in quantum materials~\cite{Wen:2019wz, PhysRevLett.100.156401}. Recent proposals have focused on charge-ordered phases in materials with pre-existing non-trivial topology~\cite{Chiu:2023aa,Shi:2021vo, Mitsuishi:2020aa}, but it is unclear whether or not topological states can emerge directly from the CDW formation itself.

Among the 2D TMDs, monolayer NbSe$_2$ is especially well-suited for exploring correlation-driven topological states due to the presence of a unique combination of strong correlation effects and trivial topology in its pristine phase. Monolayer 1H-NbSe$_2$ is especially intriguing due to its pronounced 3$\times$3 triangular CDW order ($T_{CDW} \sim 145$ K)~\cite{Xi:2015aa} and relatively weak superconductivity ($T_c \sim 1.9$ K)~\cite{Ugeda:2016aa, Wang:2017aa} compared to bulk 2H-NbSe$_2$. Recent scanning tunneling microscopy (STM) experiments have uncovered unexpected CDW phases in 2H-NbSe$_2$ under in-plane tensile strain, including a 2$\times$2 triangular phase and a 4$\times$1 stripe phase~\cite{Gao:2018aa}. These findings raise the question: Can strain induce a 2$\times$2 CDW in single-layer 1H-NbSe$_2$? While first-principles calculations have clarified the structures of the 3$\times$3 triangular~\cite{Silva_Guill_n_2016, Lian:2018aa, Guster:2019aa} and 4$\times$1 stripe~\cite{Cossu:2020aa} CDW phases in 1H-NbSe$_2$, existence of the 2$\times$2 CDW has remained elusive~\cite{Cossu:2020aa}.
 
In this study, we employ first-principles modeling and analysis to elucidate the structure of the 2$\times$2 triangular CDW phase in 1H-NbSe$_2$. We use the strongly-constrained-and-appropriately-normed (SCAN) meta-generalized-gradient-approximation (meta-GGA) functionals~\cite{PhysRevLett.115.036402}, which has proven effective across a wide range of quantum materials. \cite{PatraE9188, zhang2020, Chiu:2020aa}. We identify two stable 2$\times$2 CDW phases under in-plane biaxial tensile strain, characterized by breathing mode patterns with slightly different energies. Beyond 1\% strain, these 2$\times$2 phases become energetically favorable, replacing the 3$\times$3 phases as the ground state. Surprisingly, we discover the emergence of multiple non-trivial topological characteristics within the 2$\times$2 CDW phases. Our in-depth analysis of the band characteristics and topological features highlights how topological states can arise as a direct consequence of the CDW formation.

\section{Results and discussion}

\subsection{Strain-induced instabilities}\label{1by1}

\begin{figure}[th]
\centering
\includegraphics[width=0.8\textwidth]{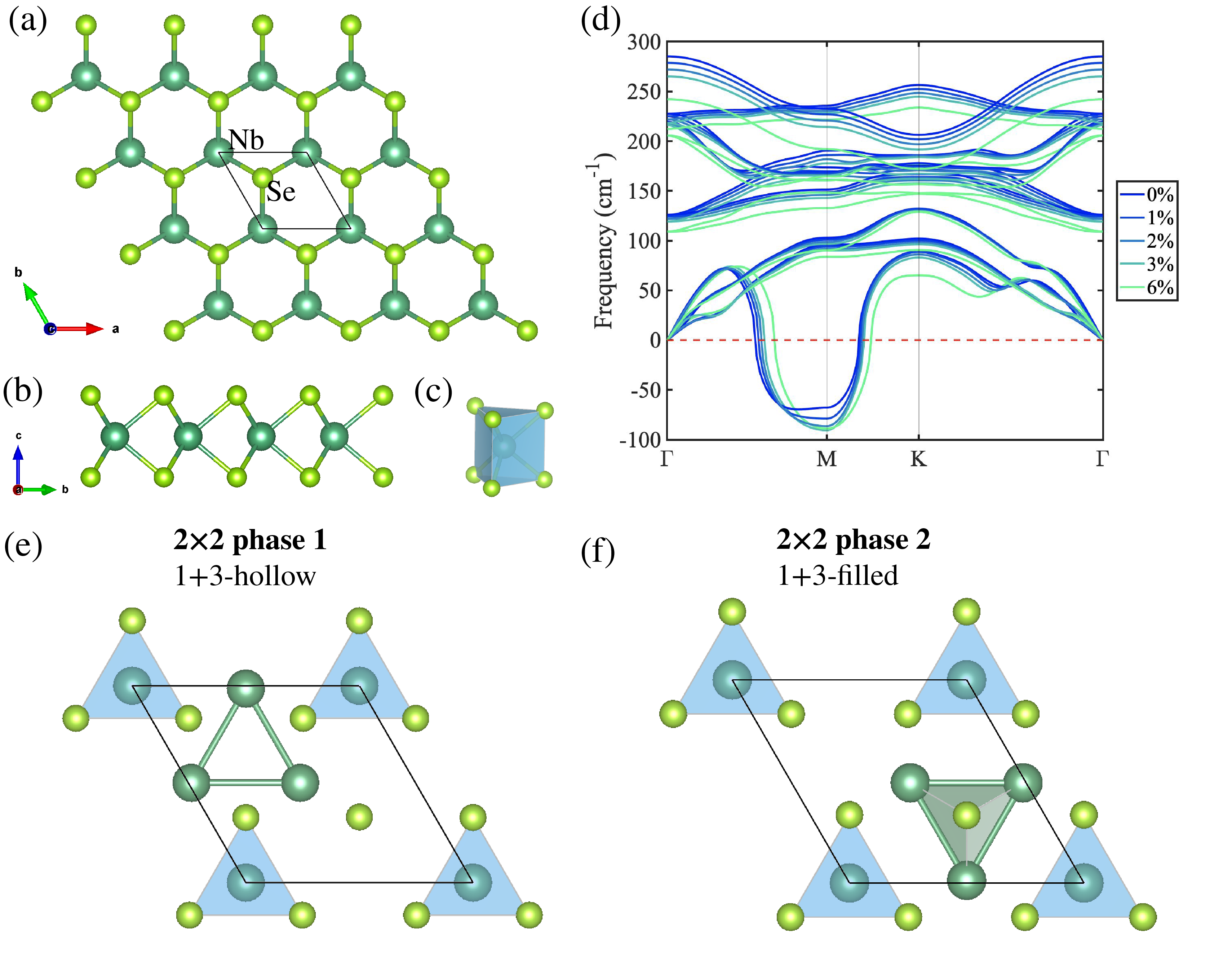}
\caption{ \textbf{Pristine and 2$\times$2 CDW phases in 1H-NbSe$_2$.} (a-c) Crystal structure of unmodulated pristine single-layer 1H-NbSe$_2$. Nb and Se atoms are shown as dark green and light green spheres. (a) Top view of the crystal structure. (b) Side view of the crystal structure. (c) The trigonal prismatic cell. (d) Phonon spectrum of unmodulated 1H-NbSe$_2$ for various in-plane biaxial strain values; results for strains up to 10\% are presented in Supplementary Fig. 1. (e) Crystal structure of the 2$\times$2 CDW Phase 1, labeled as 1+3-hollow. (f) Crystal structure of the 2$\times$2 CDW Phase 2, labeled as 1+3-filled. Green lines represent the shortened length between two adjacent Nb atoms in the CDW phases.}
\label{fig1}
\end{figure}

Bulk 2H-NbSe$_2$ is a van der Waals material with centrosymmetric space group $P6_{3}/mmc$ (No. 194) containing inversion symmetry. In contrast, the single-layer 1H-NbSe$_2$ exhibits a non-centrosymmetric space group $P\bar{6}m2$ (No. 187) with an additional out-of-plane mirror symmetry $M_z$ but lacks inversion symmetry. Fig.~\ref{fig1}(a) shows the top view of the hexagonal structure of 1H-NbSe$_2$, and the Nb layer that acts as a mirror plane for the $M_z$, which is sandwiched between the two Se layers (Fig.~\ref{fig1}(b)). Each Nb atom resides within a trigonal prismatic cell formed by its six closest Se neighbors (Fig.~\ref{fig1}(c)). Our calculations show that Nb atoms form a perfect hexagonal close-packed structure, with the shortest Nb-Nb separation being $a_0$ = 3.485 \AA.

The pristine single-layer 1H-NbSe$_2$ exhibits instability below the CDW critical temperature~\cite{Xi:2015aa, PhysRevLett.122.016403}. Our calculations identify a prominent soft-phonon mode near $\bm{q}_{CDW} \sim$ 2/3 $\Gamma$M, indicating a pronounced structural instability associated with CDW formation within a 3$\times$3 supercell, consistent with experimental observations~\cite{Xi:2015aa, PhysRevLett.122.016403}. Fig.~\ref{fig1}(d) and Supplementary Material S.1 show that as the in-plane biaxial tensile strains increase, the soft-phonon modes shift from  $\bm{q}_{CDW} \sim$ 2/3 $\Gamma$M to $\bm{q}_{CDW}  \sim$  $\Gamma$M, which implies a strain-induced structural phase transition from a 3$\times$3 to a 2$\times$2 CDW phase. Moreover, the strain-dependent soft phonon mode reveals competition between the 3$\times$3 and 2$\times$2 CDW phases for strains between 0\%-1\%. This competition is evident from the flattening of the dispersion of the soft-phonon mode between the 2/3 $\Gamma$M and the M points, which indicates that the deepest parts of the soft-phonon mode corresponding to these two $q$-vectors are competing. The broad range of $q$-vectors which simultaneously soften is consistent with a phonon-entropy dominated transition~\cite{PhysRevB.16.643}. When the strain exceeds 1\%, the deepest soft-phonon mode remains pinned at the M point, so that that the 2$\times$2 CDW becomes dominant for strains greater than 1\%.

\subsection{2$\times$2 CDW breathing mode}\label{2by2}
To reveal the 2$\times$2 CDW structure of 1H-NbSe$_2$, we conducted structural optimizations within a 2$\times$2 supercell under experimentally practical in-plane biaxial tensile strains (0\%-10\%)~\cite{Rold_n_2015}. Initial displacements that preserve the $C_3$ symmetry were applied to Nb atoms at the supercell's edges and center (triangular Nb trimer), while corner Nb atoms remained unchanged. Depending on the initial direction of these displacements, two distinct CDW phases emerged once all atomic positions were relaxed within the 2$\times$2 supercell.  Like the 3$\times$3 phases (Supplementary Material S.2.)~\cite{Lian:2018aa}, we denote the two 2$\times$2 CDWs in Figs.~\ref{fig1}(e) and (f) as Phase 1 (labeled 1+3-hollow) and Phase 2, (labeled 1+3-filled). These phases involve periodic lattice distortions (PLDs), see Supplementary Material S.3 for details, which correspond to two different distortions of a single soft Nb breathing-mode phonon that are 180$^\circ$ out of phase. Notably, both the 2$\times$2 CDW structures retain the same space group $P\bar{6}m2$ (No. 187) as the pristine structure, with the Nb atomic layer serving as a mirror plane for $M_z$. Also, our SCAN calculations show that both our 2$\times$2 CDWs converge to a nonmagnetic ground state, which is consistent with previous studies on the 3$\times$3 CDW, where the formation of the CDW is found to suppress the magnetic instability.~\cite{PhysRevB.97.081101}

\begin{figure*}[th!]
\centering
\includegraphics[width=0.9\textwidth]{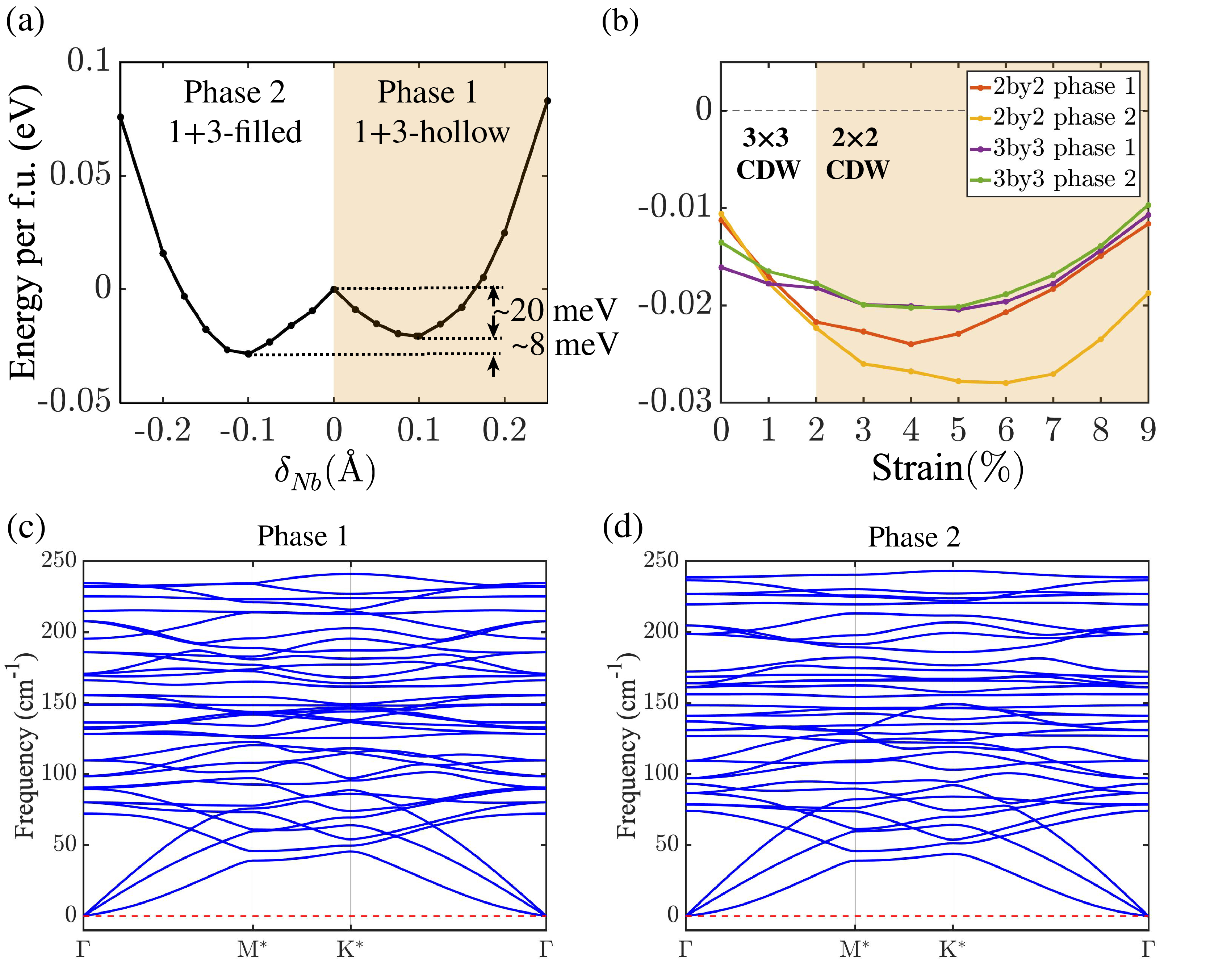}
\caption{ \textbf{Structural stability of 2$\times$2 CDW breathing mode.} (a) Energy per f.u. of the 2$\times$2 supercell under 6\% strain as a function of Nb trimer distortion $\delta_{Nb}$. (b)  Differences in total energy per f.u. relative to the energy of the pristine phase as a function of strain. The 2$\times$2 CDW phase becomes the ground state for strains greater than 1\%. (c) Phonon spectrum of Phase 1 of the 2$\times$2 CDW. (d) Phonon spectrum of Phase 2 of the 2$\times$2 CDW. }
\label{fig:2by2}
\end{figure*}

To assess structural stability, we analyze the local energy landscape by manually adjusting in-plane Nb (Nb trimer) distortions, denoted as $\delta_{Nb}$, while relaxing all other atoms. Notably, the $\delta_{Nb}$ values for each Nb within the trimer are identical, so that the $C_3$ symmetry is preserved. Fig.~\ref{fig:2by2}(a) shows the total energy per formula unit (f.u.) for different $\delta_{Nb}$ values, under 6\% strain, revealing a local maximum at $\delta_{Nb}=0$ (pristine phase) and two local minima (Phase 1 and Phase 2). The asymmetric double-well, reflecting the absence of inversion symmetry in 1H-NbSe$_2$, allows CDW transitions from the pristine phase to both CDW phases without an energy barrier.  Despite Phase 2 being slightly lower in energy, both phases are thus equally accessible during the initial CDW transition from the pristine structure. However, transitions between the energy wells of Phase 1 and Phase 2 would require a minimum energy of approximately 20 meV (equivalent to about 240K), which is higher than the CDW transition temperature (145K). Therefore, we expect the two 2$\times$2 CDW phases to coexist in strained 1H-NbSe$_2$, potentially forming a long-range landscape similar to that observed in the 3$\times$3 phases \cite{PhysRevLett.122.016403}. 

To gain deeper insight into the competition between the 2$\times$2 and 3$\times$3 CDWs, we compare their energies with two different 3$\times$3 CDW phases~\cite{Lian:2018aa, Guster:2019aa} and the pristine phase. Fig.~\ref{fig:2by2}(b) presents the energy differences per formula unit relative to the pristine phase as a function of strain, with the energy of the pristine phase at each strain normalized to zero. Notably, the 3$\times$3 CDW dominates for 0\%-1\% strain, but between 1\% to 10\% strain, both 2$\times$2 CDW phases exhibit lower energies than the 3$\times$3 CDW and the pristine phase. This strain-driven CDW phase transition from 3$\times$3 to 2$\times$2 is consistent with the shift of soft phonon modes in Fig.~\ref{fig1}(d). We expect that the commonly used techniques like chemical vapor deposition (CVD) \cite{Wang:2017aa} and molecular beam epitaxy (MBE) \cite{Ugeda:2016aa, Hotta:2016aa} for fabricating monolayer 1H-NbSe$_{2}$, combined with strain-engineering methods such as substrate-induced strain or mechanical manipulation \cite{Ahn:2017aa, Dreher:2021aa,Chen:2020aa} would allow experimental realization of our predicted 2$\times$2 CDW phases. Stability of our two 2$\times$2 CDWs is further supported by the absence of soft modes in the phonon spectrum, see Figs.~\ref{fig:2by2}(c) and (d).

\subsection{CDW-Induced Electronic Modifications} \label{bands}

\begin{figure}
\centering
\includegraphics[width=0.9\textwidth]{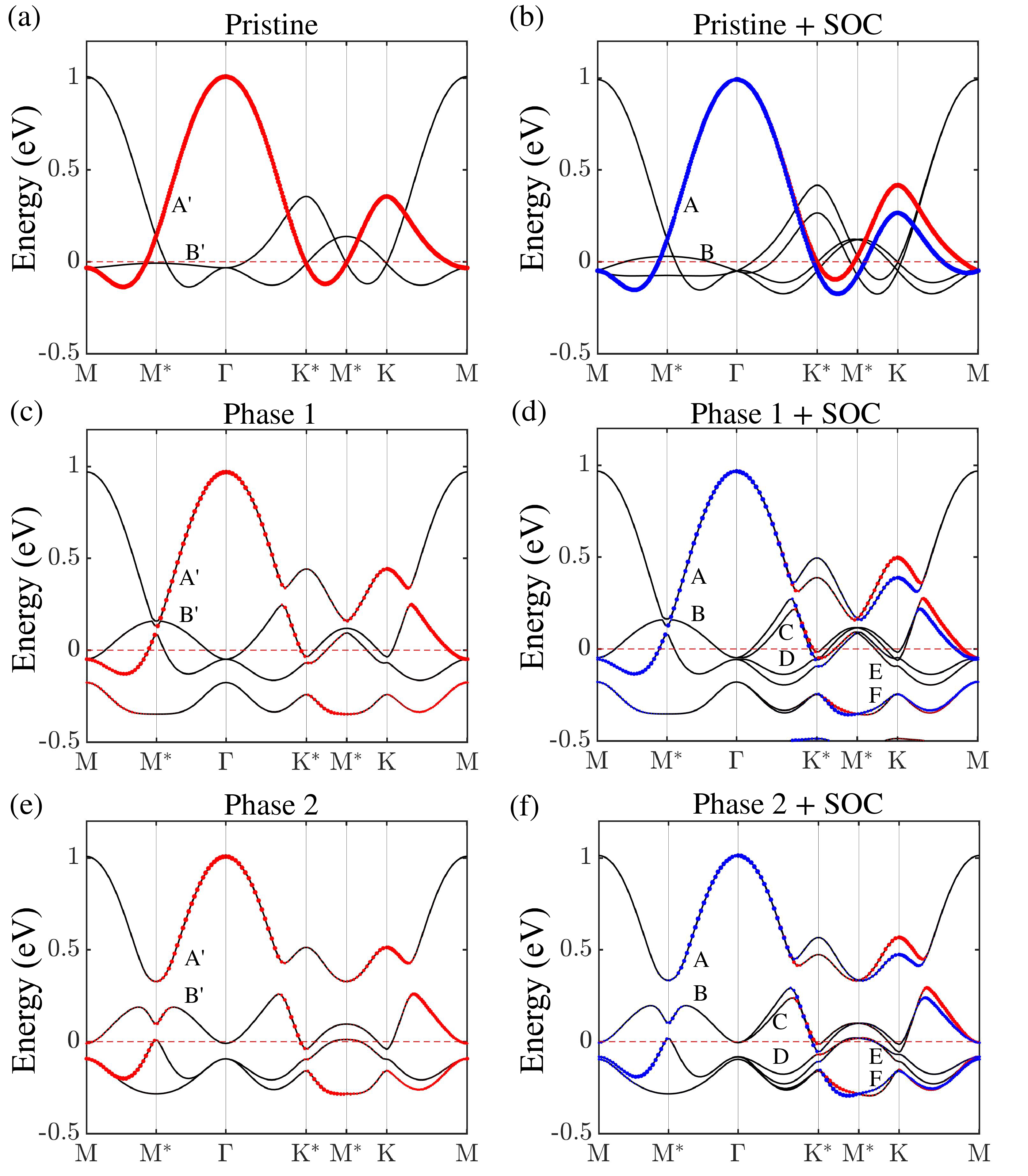}
\caption{ \textbf{Band structure of the pristine and 2$\times$2 CDW phases.} Folded and unfolded electronic band structures are shown without and with SOC. (a) and (b) represent the pristine phase. (c) and (d) correspond to Phase 1, and (e) and (f) to Phase 2 of the 2$\times$2 CDW. Black curves denote the folded bands. Unfolded spectral weights for spin-up and spin-down states are indicated by the size of the red and blue markers, respectively. The Fermi energy (red dashed line) is set to 0. The bands discussed in the text are labeled A to F. In our topological index calculations, bands B, D, and F are considered valence bands corresponding to the band gaps between A/B, C/D, and E/F, respectively.
}
\label{fig:bands}
\end{figure}

To explore the electronic structures of the 2$\times$2 CDW phases, we use 6\% strain as an representative case. At 6\% strain, the pristine structure exhibits the deepest soft-phonon mode at the M point and the Phase 2 reaches its lowest energy configuration. In Fig.~\ref{fig:bands}, black curves depict the folded band structure within the 2$\times$2 supercell BZ. The unfolded spin-up (red) and spin-down (blue) bands illustrate how the CDW modifies the band structure relative to the pristine phase within the 1$\times$1 Brillouin zone (BZ). Note that the first BZ of the 2$\times$2 supercell  is folded from the BZ of the 1$\times$1 unit cell, with the high-symmetry points illustrated in Fig.~\ref{fig4}(c). 

By comparing the band structures of CDW phases with the pristine phase with and without spin-orbit coupling (SOC) in Figs.~\ref{fig:bands}(a) and (b), we can see significant changes induced by the 2$\times$2 CDWs. In Phase 1, without the SOC (Fig.~\ref{fig:bands}(c)) the CDW opens a gap at the folding point of band A' at M$^\ast$, and elevates the relatively flat B' band above this CDW gap, resulting in a band inversion feature. Importantly, SOC alone does not raise band B above the crossing points of band A in the folded band structure of the pristine phase in Figs.~\ref{fig4}(a) and (b), which provide zoomed-in views along the $\Gamma$-M line from Figs. \ref{fig:bands}(a) and (b). These results imply that the band inversion between bands A' and B' around M$^\ast$ in Phase 1 is a direct consequence of the CDW. When the SOC is introduced in Phase 1, it opens a fundamental gap between bands A and B (Figs.~\ref{fig4}(e)). In contrast, in Phase 2 without SOC (Fig.~\ref{fig:bands}(e)), the 2$\times$2 CDW not only induces the band inversion but it also creates a full gap between bands A' and B'. This further indicates that the band inversion features are driven by the CDW rather than the SOC.


\subsection{Emergence of Topological States in CDW Phases} \label{Topo}

\begin{figure*}
\centering
\includegraphics[width=0.9\textwidth]{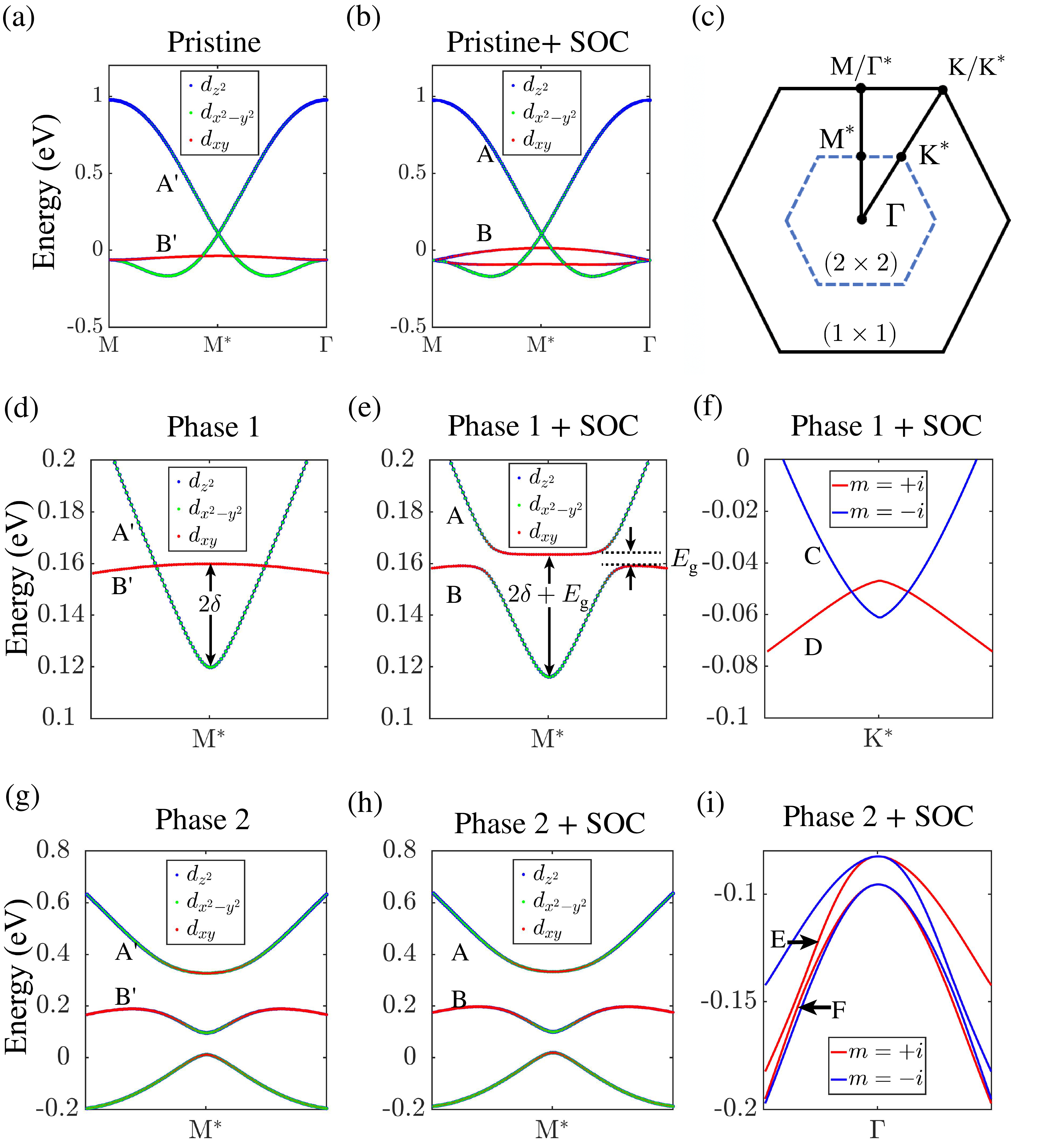}
\caption{\textbf{Topological band structure in 2$\times$2 CDW phases.} Energy dispersions along the $\Gamma$-M$^\ast$-$\Gamma$ path without/with SOC for the pristine phase (a, b), for Phase 1 (d, e), and for Phase 2 (g, h). The dominant Nb $d_{xy}$, $d_{z^2}$, and $d_{x^2-y^2}$ orbital contributions are illustrated by the size of the red, blue, and green markers, respectively. (c) First BZs of the monolayer in the 1$\times$1 pritine phase (black solid lines) and the 2$\times$2 CDW phase (blue dashed lines) with three high-symmetry points, $\Gamma$, M, and K. High-symmetry points of the 2$\times$2 BZ are marked with a star ($^\ast$). (d, e) Only bands A and B around M$^\ast$ along the $\Gamma$-M$^\ast$ path of Phase 1 are shown. $2\delta$ represents the CDW-induced inverted gap, while $E_g$ denotes the SOC-induced fundamental gap. (f) Energy dispersion of Phase 1 with SOC along the $\Gamma$-K$^{\ast}$ path between bands C and D, indicating the formation of a nodal ring around K$^{\ast}$. (g, h) Only the energy dispersion around M$^\ast$ along the $\Gamma$-M$^\ast$ path of Phase 2 is shown. (i) Band structure of Phase 2 with SOC along the $\Gamma$-K$^{\ast}$ path, showing a full gap opening-up between bands E and F.
In (f) and (i), the eigenvalues $m=+i$/$m=-i$ of the out-of-plane mirror symmetry $M_z$ are represented by red/blue curves.}
\label{fig4}
\end{figure*}

The band inversion resulting from 2$\times$2 CDWs suggests the possible presence of a non-trivial charge-ordered topology. Due to the presence of time-reversal symmetry (TRS) in both 2$\times$2 phases, we first examine the TRS protected $\mathbb{Z}_2$ index within the CDW induced gaps between bands A and B. The $\mathbb{Z}_2$ index calculation reveals a non-trivial topology ($\mathbb{Z}_2=1$) for Phase 1, while Phase 2 remains topologically trivial ($\mathbb{Z}_2=0$), as detailed in the Supplementary Materials S.5. An analysis of orbital characters provides an explanation for why the $\mathbb{Z}_2$ index is non-trivial for Phase 1 but not for Phase 2. In Phase 1, without the SOC (Fig.~\ref{fig4}(d)), the CDW-induced band inversion creates an inverted gap of $2\delta$ at the M$^\ast$ point. Due to the mirror and time-reversal symmetries, this inversion is accompanied by the appearance of six Dirac points near M$^\ast$ in the 2$\times$2 BZ (Supplementary Materials S.4.). The introduction of SOC results in the opening of a fundamental gap $E_g$ at these Dirac points (Fig.~\ref{fig4}(e)), featuring a non-trivial $\mathbb{Z}_2$ topological feature. In contrast, Phase 2 exhibits a full gap between bands A and B (Fig.~\ref{fig4}(h)),  even without the SOC (Fig.~\ref{fig4}(g)), and the double-band inversion of $d_{xy}$  bands at M$^{\ast}$ results in a trivial topology between these bands.  


The mirror symmetry $M_z$, which is present throughout the 2D BZ, allows us to categorize all bands based on their mirror eigenvalues $m=\pm i$. Thus, we also examined the mirror Chern number for the full CDW gaps. In Phase 1, when considering occupied bands up to band B (Fig.~\ref{fig4}(e)), we find a mirror Chern number $C_m=3$. In Phase 2, counting occupied bands up to band F yields $C_m=2$. The band structure between bands E and F around $\Gamma$ along K$^\ast$-$\Gamma$-K$^\ast$ path (Fig.~\ref{fig4}(i)) demonstrates that the inverted mirror eigenvalue of band F causes bands E and F to have the same sign of mirror eigenvalue along this path, leading to the non-trivial mirror Chern number. These high mirror Chern numbers indicate that both 2$\times$2 CDW phases host topologically non-trivial states within their CDW gaps.  

We also find nodal rings around each K$^{\ast}$ between bands C and D in Phase 1, as shown in Fig.~\ref{fig4}(f) and the Supplementary Fig.~4. The topology of these rings is characterized by the zero-dimensional enclosing manifold, with topological index $\zeta_0 = N_{P_2,i} - N_{P_1,i} = 26 - 25 = 1$, indicating a mirror-protected topological nodal line. Intriguingly, in Phase 2, although there is no nodal ring between bands C and D, the fully gapped bands C and D, along with the presence of a double band inversion at M$^{\ast}$, suggest the emergence of higher-order topology. 

We emphasize that all the topological features discussed above arise from correlation effects acting on the initially topologically trivial band crossing the Fermi level, underscoring the significance of the CDW-induced topology. In contrast, the 3$\times$3 CDWs in monolayer 1H-NbSe$_2$ do not open a global band gap on the topologically trivial band of the pristine phase\cite{Silva_Guill_n_2016, Lian:2018aa}, and thus cannot induce topological features that require a global band gap to be well-defined. This distinction highlights the key role of the strain-induced 2$\times$2 CDW in generating non-trivial topological states.


\begin{figure}
\centering
\includegraphics[width=0.9\textwidth]{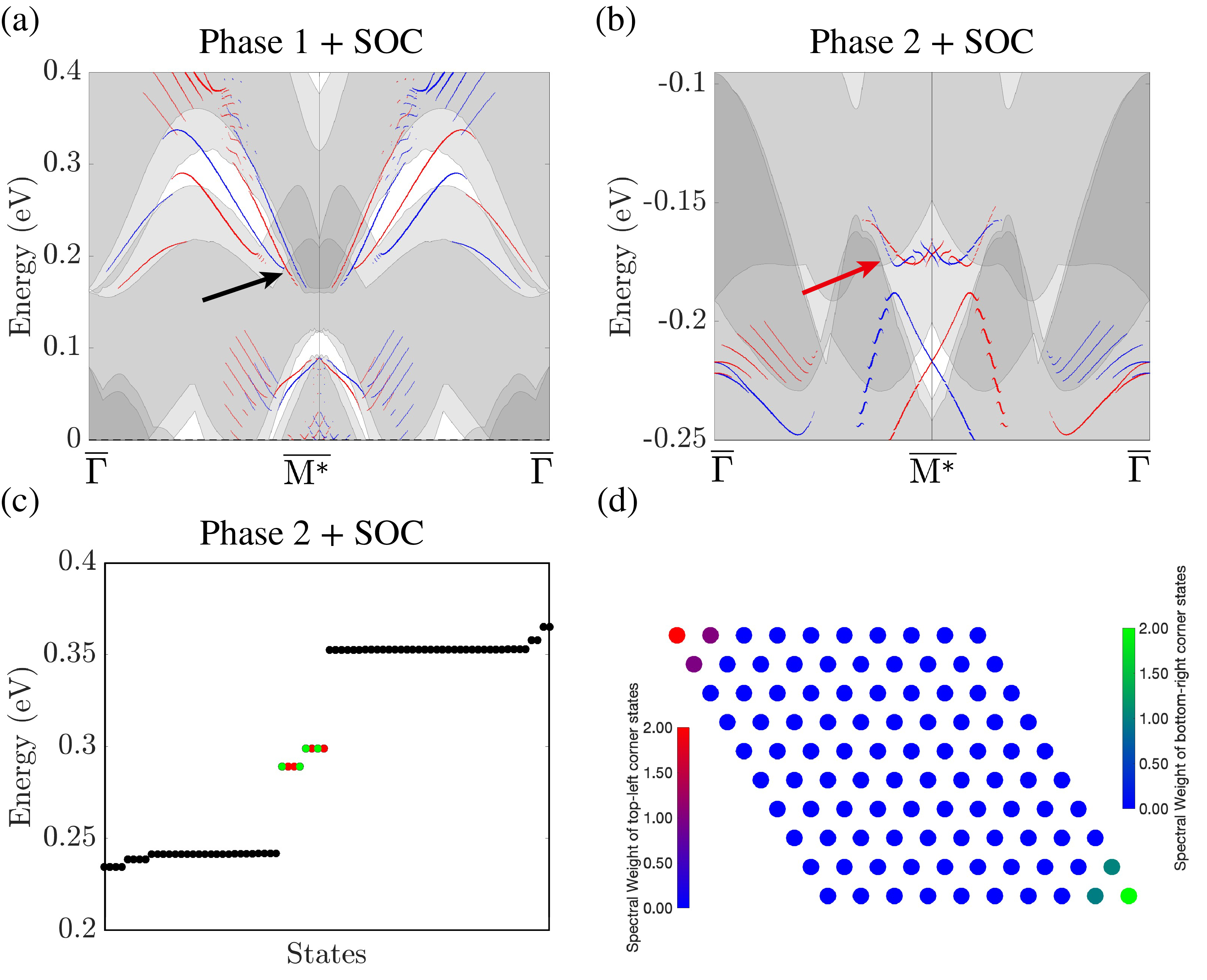}
\caption{\textbf{Non-trivial topology in the CDW gaps.} 
(a) Edge band structure of Phase 1 along the $\overline{\Gamma}-\overline{\text{M}^\ast}-\overline{\Gamma}$ of the [100] ribbon with 30 layers, showing three edge states (black arrow) corresponding to $C_m=3$ and $\mathbb{Z}_2=1$ in the CDW gap between bands A and B. Grey structures denote bulk projections along the ribbon direction, while blue and red markers indicate edge state spin-polarization. Marker size corresponds to the spectral weight of layer 1 with a lower bound cutoff of 0.08. (b) Edge band structure of Phase 2 along $\overline{\Gamma}-\overline{\text{M}^\ast}-\overline{\Gamma}$ of the [010] ribbon with 30 layers, revealing two discernible edge states (red arrow) corresponding to $C_m=2$ in the CDW gap between band E and band F of Phase 2. (c) Energy spectrum of a $10\times10$ nanodisk for Phase 2, focusing on the energy range between bands A and B. The black markers indicate edge states, while red and green markers indicate the top-left and bottom-right corner states, respectively. (d) Real-space distribution of the eight corner states within the gap between bands A and B in the nanodisk. Color gradient indicates the spectral weights of the corner states: red-to-blue for the top-left corner states and green-to-blue for the bottom-right corner states, with red and green representing the highest intensity and blue representing the lowest.}

\label{fig:topo}
\end{figure}

\section{Topological Edge and Corner States} 
The existence of nontrivial topology in 2D materials can also be identified through their edge states. In Phase 1, the edge band structure of the [100] ribbon is illustrated in Fig.~\ref{fig:topo}(a), with further details available in the Supplementary Material S.6. Although the bulk band gap around M$^{\ast}$ is substantially covered by the projections of bulk bands (the grey region), we can still clearly observe three distinct edge states (black arrow in Fig.~\ref{fig:topo}(a)) that emerge between bands A and B of Phase 1, at around 0.16eV above the Fermi level. These three edge states correspond to $C_m=3$ and $\mathbb{Z}_2=1$, where $C_m=3$ indicates the existence of three edge states connecting bands A and B, and $\mathbb{Z}_2=1$ requires that an odd number of edge states bridge this gap, confirming the nontrivial topology.

In Phase 2, we present the edge band structure of a [010] ribbon, and focus on the connection between bands E and F. Despite substantial overlap with bulk band projections within the band gap, two distinct edge states can still be identified (red arrow in Fig.~\ref{fig:topo}(b); further details in the SM). These two edge states, bridging bands E and F, signify the presence of two edge states corresponding to $C_m=2$ within this CDW-induced gap.  We further elucidate the higher-order topological features in Phase 2 in Figs.~\ref{fig:topo}(c) and (d). Fig.~\ref{fig:topo}(c) shows the energy spectrum of a $10 \times 10$ nanodisk, focusing on the energy range between bands A and B, where the black markers represent edge states within this CDW gap.  Notably, in Fig.~\ref{fig:topo}(c), we observe eight corner states distributed across two energy levels emerging alongside the gapped edge states, where the red markers represent corner states localized at the top-left corner and the green markers indicate those at the bottom-right corner, with each corner hosting two spin-up and two spin-down states. Fig.~\ref{fig:topo}(d) shows the real-space distribution of these corner states, illustrating their localization at the top-left and bottom-right corners of the nanodisk. The total spectral weight sums to 4 at each corner, indicating the presence of four corner states, see Supplementary Material S.7. These localized corner states underscore the higher-order topological character induced by the CDW in Phase 2.

\section{Conclusion}\label{conclusion}

In summary, our first-principles modeling and analysis reveal the presence of two distinct stable structures of strain-induced 2$\times$2 CDW phases in 1H-NbSe$_{2}$. Our findings show that a modest biaxial tensile strain (>1\%) drives a transition from the 3$\times$3 to 2$\times$2 CDW phases, a shift supported by a soft phonon-mode and the inversion in energy hierarchy.  We expect this strain-induced transition to be amenable to experimental realization. Our analysis suggests the potential coexistence of both 2$\times$2 phases in strained 1H-NbSe$_{2}$; further STM studies in this connection would be interesting. 

We also uncover the coexistence of multiple emergent topological states near the Fermi energy within the two strain-induced 2$\times$2 CDW phases. Phase 1 exhibits a distinctive combination of topological features, including a non-trivial $\mathbb{Z}_2$ invariant, a high mirror Chern number ($C_m=3$), and mirror-protected topological nodal rings. In contrast, Phase 2 is characterized by a high mirror Chern number ($C_m=2$) and the presence of higher-order topological states. These topological states are rigorously verified not only by their corresponding topological indices but also by the presence of robust edge and corner states, demonstrating their intrinsic connection to the periodic lattice distortions and symmetry-breaking induced by the CDW.

Our prediction of CDW-induced topological states in 1H-NbSe$_{2}$, along with the identification of competing CDW phases, contributes to the understanding of emergent topology in strongly correlated materials. These correlation‑driven topological states attainable with modest strain may be applied in electronics, spintronics, and other quantum technologies. Given the structural and electronic similarities across the TMDs, our findings on strain-induced 2$\times$2 CDW phases in 1H-NbSe$_{2}$ likely extend to other monolayer TMDs, such as 1H-TaS$_{2}$\cite{Hall:2019aa}, 1H-TaSe$_{2}$\cite{Shi:2018aa}, and 1H-NbS$_{2}$\cite{Knispel:2024aa}, which exhibit similar 3$\times$3 CDW behavior and comparable band structures, possible sensitivities to microscopic factors such as doping\cite{Knispel:2024aa} notwithstanding.

\section{Methods}\label{Methods}
The lattice dynamics calculations of the unmodulated 1$\times$1 structure were carried out within the framework of density functional perturbation theory (DFPT)~\cite{PhysRevLett.58.1861} using norm-conserving pseudo-potentials~\cite{PhysRevB.43.1993} as implemented in the Quantum Espresso package~\cite{Giannozzi_2009}. Exchange-correlation effects were treated within the local density approximation (LDA)~\cite{PhysRev.140.A1133} with the Perdew-Zunger parametrization~\cite{PhysRevB.23.5048}. We have fully optimized both the ionic positions and lattice parameters until the residual forces on each ion were less than $10^{-4}$ Ry and zero-stress tensors are obtained. The monolayer is obtained from the relaxed bulk structure by setting a vacuum of 14 $\AA$ along the z-direction to eliminate interaction with replica images. This was followed with another ionic relaxation step to ensure structural stability. The in-plane biaxial tensile strain was simulated by increasing the lattice parameter based on the relaxed monolayer structure. For the strained 1$\times$1 structure, we only relaxed the atomic positions by fixing the lattice parameter and unit cell shape. For the pristine phase, phonon dispersion was obtained by Fourier interpolation of the dynamical matrices computed on a $24\times24\times1$ $k$-mesh and $12\times12\times1$ $q-$mesh. For the 2$\times$2 CDW phase, the Brillouin zones for the electronic and vibrational calculations were sampled using $9\times9\times1$ and $2\times2\times1$ meshes, respectively. 

Lattice relaxation of the modulated structures and the electronic structure calculations were all carried out within the DFT framework with the projector augmented wave (PAW) method using the Vienna \textit{ab initio}  Simulation Package (VASP)~\cite{PhysRevB.59.1758,PhysRev.136.B864,PhysRev.140.A1133,PhysRevB.54.11169,PhysRevLett.77.3865}. We used the strongly-constrained-and-appropriately-normed (SCAN) meta-generalized-gradient-approximation (meta-GGA) functional with the Perdew-Burke-Ernzerhof (PBE) parametrization~\cite{PhysRevLett.115.036402} to include exchange-correlation effects. An energy cut-off of 400\,eV was used for the plane-wave-basis set. To keep approximately the same distance between the sampling $k$-points, the $\Gamma$-centered $21\times21\times1$, $11\times11\times1$ and $7\times7\times1$ $k$-meshes was employed accordingly to sample the 2D Brillouin zones (BZ) of the $1\times1$, $2\times2$ and $3\times3$ monolayer structures, respectively.  The ionic relaxation of modulated 2$\times$2 and 3$\times$3 structures was converged until residual forces were smaller than $10^{-3}$ eV\AA$^{-1}$ per atom and energy tolerance of $10^{-4} $eV per unit cell. The first order Methfessel-Paxton smearing method with smearing width 0.1eV was used in the ionic relaxation process, the tetrahedron method with smearing width 0.1eV was used in the self-consistent run of electronic structure, and the Gaussian smearing with smearing width 0.1eV was used to calculate the electronic band structures along the high-symmetry linen. SOC was not included in the ionic relaxation process, but it was included self-consistently in the electronic structure calculations. The topological analysis was performed by employing a real-space tight-binding model Hamiltonian, which was obtained by using the VASP2WANNIER90 interface~\cite{PhysRevB.56.12847}. Nb \textit{d} and Se \textit{p} states were included in generating Wannier functions. All crystal structures are visualized using the VESTA~\cite{Momma:db5098} package.

To analyze the topological properties of the 2$\times$2 CDW phases in 1H-NbSe$_2$, we employed several computational methods. The $\mathbb{Z}_2$ index was computed using the Wilson loop method ~\cite{PhysRevB.83.235401} given the absence of inversion symmetry in the 1H structure.  For the mirror Chern number calculations, we calculate the Wilson loop with mirror eigenvalues $m=\pm i$ separation. The accumulated geometric phase difference can be expressed as $\Theta^{m}\equiv\int \dot{\vartheta }^m_{k_y}dk_{y}$ along the principle axis $k_y$, where $\dot{\vartheta }^m_{k_y} \equiv \sum^{occ}_{n=1} d\vartheta^{{}m}_{k_{y},n} /dk_{y}$ with $\vartheta$ as the Berry phases, $n$ as the band index and $occ$ as the number of occupied bands. The first Chern class of the sub Hilbert space with each mirror eigenvalue $m$ is identified as $C_{\pm i}=\Theta^{\pm i}/{2\pi }$, and the mirror Chern number was calculated as $C_m=\frac{1}{2 }(C_{+i}-C_{-i})$. The topology of the nodal ring is confirmed by the zero-dimensional enclosing manifold~\cite{Fang_2016}. By counting the number of occupied bands with $M_z$ eigenvalue $m=i$ up to the energy of the nodal ring at a point inside (outside) of the nodal ring denoted as $p_1$ ($p_2$), the $\mathbb{Z}$ class topological invariance is defined as $\zeta_0=N_{P_2,i}-N_{P_1,i}$.

\begin{acknowledgement}
The authors thank Chao-Sheng Lian, for providing their 3$\times$3 CDW structures~\cite{Lian:2018aa}. The work at Northeastern University was supported by the National Science Foundation through the Expand-QISE award NSF-OMA-2329067 and benefited from the resources of Northeastern University’s Advanced Scientific Computation Center, the Discovery Cluster, the Massachusetts Technology Collaborative award MTC-22032, and the Quantum Materials and Sensing Institute. H.L. acknowledges the support by Academia Sinica in Taiwan under grant number AS-iMATE-113-15. S.M. and A.A. gratefully acknowledge the HPC facility at Indian Institute of Technology Kanpur, for computational resources. J.N.  benefited from resources of the Tampere Center for Scientific Computing, TCSC. T.H.'s research is supported by the ITU-BAP project TDK-2018-41181 and numerical calculations were performed at TUBITAK ULAKBIM, High Performance and Grid Computing Center TRUBA. 
T.-R.C. was supported by National Science and Technology Council (NSTC) in Taiwan (Program No. MOST111-2628-M-006-003-MY3 and NSTC113-2124-M-006-009-MY3), National Cheng Kung University (NCKU), Taiwan, and National Center for Theoretical Sciences, Taiwan. This research was supported, in part, by the Higher Education Sprout Project, Ministry of Education to the Headquarters of University Advancement at NCKU. T.-R.C. thanks the National Center for Highperformance Computing (NCHC) of National Applied Research Laboratories (NARLabs) in Taiwan for providing computational and storage resources.
The work at TIFR Mumbai was supported by the Department of Atomic Energy of the Government of India under Project No. 12- R\&D-TFR-5.10-0100 and benefited from the computational resources of TIFR Mumbai.
\end{acknowledgement}

\begin{suppinfo}

The Supporting Information is available free of charge.
\begin{itemize}
  \item Phonon spectrum of the pristine structure; 3$\times $3 CDW crystal structure of 1H-NbSe$_2$; periodic lattice distortion in 2×2 CDW phases; topological band features; Wilson loop calculations; calculations of topological edge states; topological corner states; hybrid functional calculations
 
\end{itemize}

\end{suppinfo}

\bibliography{NbSe2}

\end{document}